\documentclass[12pt]{article}
\usepackage{amssymb,amsmath,epsfig}
\begin{document} \parindent=0pt
\parskip=6pt \rm

\vspace{0.5cm}
\begin{center}
 {\bf \large  Comment on: ``First-order phase transitions of  \\ type-I
superconducting films'' [Phys. Lett. A 322 (2004) 111]}

\vspace{0.5cm}

{\bf  D. V. Shopova$^{\ast}$ and D. I. Uzunov$^{\ast,
\dag}\footnote{Corresponding author: uzun@issp.bas.bg}$}

\vspace{0.2cm}

{\em  CP Laboratory, G. Nadjakov
Institute of Solid State Physics,\\
Bulgarian Academy of Sciences, BG-1784
Sofia, Bulgaria}

$^{\dag}$ {\em Max-Planck-Institut f\"ur Physik komplexer Systeme,\\
N\"othnitzer Str. 38, 01187 Dresden, Germany }
\end{center}



\vspace{0.5cm}

The theory of fluctuation induced first order phase transition
from normal to uniform (Meissner) superconducting state in a zero
external magnetic field was developed for thin films of type I
superconductors in a series of papers~\cite{Folk:2001,
Rahola:2001, Shopova:2002, Shopova1:2003, Shopova2:2003,
Shopova3:2003, Shopova4:2003}. The  same problem was considered in
a recent theoretical investigation~\cite{Abreu:2004}. The authors
of the paper~\cite{Abreu:2004} derived an erroneous effective free
energy of superconducting slabs (films) and this error led them to
entirely wrong conclusions. Here we shall make clear the genesis
of this mistake by using the notations in~\cite{Abreu:2004}.

At a certain stage of derivation of
the effective free energy $\cal{F}(\phi)$ as a function of the
modulus $(\phi \equiv |\psi|)$ of the superconducting order parameter
$\psi$, the authors of Ref.~\cite{Abreu:2004} have obtained the
following result
\begin{equation}
\label{eq1}
\frac{d{\cal{F}}(\phi)}{d\phi} = m_0^2\phi + \frac{\lambda}{2}\phi^3 +
(d-1)e^2\phi I(\phi),
\end{equation}
where $m_0^2 = (T/T_0 - 1)$ and $\lambda$ are well known Landau
parameters of the $\phi^4-$theory of second order phase
transitions~\cite{Uzunov:1993}. The spatial dimensionality is
denoted by $d$, $e = |e|$ is the electron charge, and
\begin{equation}
\label{eq2}
I(\phi) = (2\pi)^{-\frac{d}{2}}\left[2^{-\frac{d}{2}}\Gamma\left (1-
\frac{d}{2} \right)
\left(e\phi\right)^{d-2} + 2\sum_{n=1}^{\infty}\left(\frac{e\phi}{nL}\right)
^{\frac{d}{2}-1} K_{\frac{d}{2}-1}\left(e\phi Ln\right)\right]
\end{equation}
(cf Eqs.~(5) and (11) in~\cite{Abreu:2004}). In (2), $K_{\nu}(z)$
is the MacDonald function~\cite{Prudnikov:1986}. All quantities in
(1) and (2) are dimensionless and,  in particular lengths, such as
the film thickness $L$, are scaled by the zero-temperature
coherence length $\xi_0$ ($L \rightarrow
L/\xi_0$)~\cite{Abreu:2004}.

 The compact notation
\begin{equation}
\label{eq3} R_n^{\nu}(\phi) = \left(e\phi Ln\right)^{\nu}
K_{\nu}\left(e\phi Ln\right),
\end{equation}
will be used in the remainder of this Comment. The integral
formula
\begin{equation}
\label{eq4}
\int_{0}^{y} x^{\frac{d}{2}}K_{\frac{d}{2} - 1}(x)dx = -y^{\frac{d}{2}} K_{\frac{d}{2}}\left(y\right) +
2^{\frac{d}{2}-1}\Gamma\left(\frac{d}{2}\right),
\end{equation}
is valid for dimensionalities $d > 0$~\cite{Prudnikov:1986}, and
we shall apply the known expansion of the function $K_{\nu}(z)$
for small values of $z$ (see, e.g., ~\cite{Stegun:1972}).

In order to obtain the derivative $d{\cal{F}}/d\phi$ to the
relevant order $\phi^3$, one has to expand the second term in
r.h.s. of Eq.~(2) to order $\phi^2$. The results should be valid
for spatial dimensionalities $2 < d < 4$. The authors
of~\cite{Abreu:2004} have performed
 an expansion of the second term in r.h.s. of Eq.~(2) to order $\phi^0$
and failed to reveal a term of order $\phi^{d-2}$ that exactly
cancels the first term in r.h.s. of (2). The expansion contains
also a term of type $\phi^2$ which gives a fluctuation
contribution to the $\phi^4-$invariant in the free energy as shown
below. This means that the free energy ${\cal{F}}(\phi)$ does not
contain a $\phi^d-$term for $2 < d <4$ and, hence, this free
energy does
 not describe a first order phase transition at all.

Here we show the mentioned error in [8] by the calculation of the free energy
${\cal{F}}(\phi)$ to order $\phi^4$. From (1) and (2) we obtain
\begin{equation}
\label{eq5}
{\cal{F}} = \frac{m_0^2}{2}\phi^2 + \frac{\lambda}{8}\phi^4 + (d-1)e^2J(\phi),
\end{equation}
where the integral
\begin{equation}
\label{eq6}
J(\phi) = \int^{\phi}_0d\varphi\varphi I(\varphi),
\end{equation}
is calculated using (4). With the help of (3) $J(\phi)$ can be
written in the form
\begin{equation}
\label{eq7} J(\phi) =
(2\pi)^{-\frac{d}{2}}\left[\frac{2^{-\frac{d}{2}}}{d}\Gamma\left
(1-\frac{d}{2}\right) e^{d-2}\phi^{d} +
\frac{2^{\frac{d}{2}}\Gamma(d/2)\zeta(d)}{e^2L^d}
-\frac{2}{e^2L^d}\sum_{n=1}^{\infty}\frac{R_n^{d/2}(\phi)}{n^d}\right].
\end{equation}
 Now the expansion of $R_n^{d/2}(\phi)$
for $(e\phi Ln) \ll 1$ straightforwardly leads to the following
form of the sum in (7):
\begin{eqnarray}
\label{eq8}
\sum_{n=1}^{\infty}\frac{R_n^{d/2}}{n^d}& = &2^{\frac{d}{2}-1}
\Gamma(d/2) \\ \nonumber
&& \left[ \zeta(d)+\frac{\zeta(d-2)y^2}{2(2-d)}
+ \frac{\zeta(d-4)y^4}{8(4-d)(2-d)}
+\frac{\Gamma(-d/2)\zeta(0)y^d}{2^d\Gamma(d/2)}\right],
\end{eqnarray}
where $y = (e\phi L)$, and $\zeta(z)$ is the (Riemann) zeta
function~\cite{Stegun:1972}. Inserting the result (8) in (7) and
having in mind that $\zeta(0) = -1/2$ one easily obtains that the
first two terms in r.h.s. of (7) are cancelled by two new terms of
type $\phi^0$ and $\phi^d$ which drop from the expansion of the
third term in r.h.s. of (7). We shall not show this elementary
calculation.

Therefore, the fluctuation correction $\Delta {\cal{F}} =
(d-1)e^2J(\phi)$ to ${\cal{F}}(\phi)$ does not contain any
$\phi^d$ term; in particular, $\phi^3$ term for $d=3$. Thus, the
proof that the effective free energy, given by Eq.~(14)
in~\cite{Abreu:2004} does not follow from the Eqs.~(5) and (11) in
the same paper~\cite{Abreu:2004} is completed.

Our present analysis shows that the fluctuation contributions to
the free energy in~\cite{Abreu:2004} should be given by
\begin{equation}
\label{eq9}
\Delta{\cal{F}}(\phi) = \frac{2^{\frac{d}{2}-1}(d-1)\Gamma(d/2)}{\pi^{d/2}
(d-2)L^d}\left[\zeta(d-2)(e\phi L)^2 + \frac{\zeta(d-4)}{4(4-d)}(e\phi L)^4
\right],
\end{equation}
 For $d \rightarrow 3$ the first
term in (9) is divergent. To avoid this singularity the authors
of~\cite{Abreu:2004} performed a renormalization. In this way they
 achieved a finite negative shift ($-e^2T_0/L$)
 of the bulk characteristic temperature $T_0$. The shift
vanishes in the bulk limit $L \rightarrow \infty$, as should be,
 but the second term
in (9), which has been overlooked in~\cite{Abreu:2004}, tends to
large negative values for $L \gg 1$ and $d < 4$. When $L
\rightarrow \infty$, the magnitude of the large negative value of
the fluctuation contribution exceeds the value of the positive
parameter $\lambda$ and the superconducting phase becomes
unstable. In such a situation there is no phase transition, or,
alternatively, terms of order in $\phi$ higher than fourth should
be included in the free energy with the aim to ensure the
description of tricritical and/or other multicritical points and
another type of first order phase transitions~\cite{Uzunov:1993}.
In the present case, this scenario is impossible because all terms
of order $\phi^a$ ($a>4$) generated by (7) tend to infinity for $L
\rightarrow \infty$. This simply means that the method used in
 \cite{Abreu:2004} is unreliable.

The error in the derivation of Eq.~(14) in~\cite{Abreu:2004} totally
 invalidates
the thermodynamic analysis performed in the second part of
~\cite{Abreu:2004} and intended to describe the thermodynamic properties of
(almost) three dimensional slabs. Besides, several new incorrect points of
view are introduced in this thermodynamic analysis. For example, the correct
 equilibrium phase transition temperature has not been calculated and,
moreover, the latter has been wrongly identified with the
temperature which nullifies the coefficient in front of the
$\phi^2-$term. Let us emphasize that the temperature, at which the
coefficient of the $\phi^2-$term becomes equal to zero, is never
the phase transition temperature of a first-order phase transition
produced by a $\phi^3$-term (see, e.g., Ref.~\cite{Uzunov:1993}).
In their thermodynamic analysis the authors of~\cite{Abreu:2004}
missed
 to take advantage of the thorough consideration of three-dimensional
(bulk) superconductors presented in Ref.~\cite{Shopova4:2003}.

Finally we wish to stress that the authors~\cite{Abreu:2004} have
made
 a quite inappropriate comparison of their results
with the results by our coauthors and ourselves~\cite{Folk:2001,
Shopova1:2003, Shopova2:2003, Shopova3:2003}. Our results are
intended to describe quasi-two-dimensional films, where the
magnetic fluctuations affect the phase transition properties by an
entirely different mechanism, namely, by a free energy term of the
type $|\psi|^2\mbox{ln}|\psi|$. This low-dimensional limit cannot
be achieved by a simple analytical continuation of results
obtained for dimensionalities $2 < d < 4$.

{\bf Acknowledgments}

 DIU thanks the hospitality of
MPI-PKS-Dresden. Financial support by SCENET (Parma) and JINR (Dubna) is also
acknowledged.


\begin{thebibliography}{ll}

\bibitem{Folk:2001}
R. Folk, D.V. Shopova, and D.I. Uzunov, Phys. Lett. A 281 (2003) 197.
\bibitem
{Rahola:2001}
J.C. Rahola, J. Phys. Studies 5 (2001) 304.
\bibitem{Shopova:2002}
D.V. Shopova, T.P. Todorov, T.E. Tsvetkov, and D.I. Uzunov,
 Mod. Phys. Lett. B 16 (2002) 829.
\bibitem{Shopova1:2003}
D.V. Shopova, T.P. Todorov, and D.I. Uzunov,
 Mod. Phys. Lett. B 17 (2003) 141.
\bibitem{Shopova2:2003}
D.V. Shopova, T.P. Todorov,
J. Phys. Cond. Matter. 15 (2003) 5793; cond-mat/0305586.
\bibitem{Shopova3:2003}
D.V. Shopova, T.P. Todorov, Phys. Lett. A 314 (2003) 250.
\bibitem{Shopova4:2003}
D.V. Shopova, T.P. Todorov, J. Phys. Studies 7 (2003) 330;
arXiv: cond-mat/0306466.
\bibitem{Abreu:2004}
L.M. Abreu, A.P.C. Malbouisson, Phys. Lett. A 322 (2004) 111.
\bibitem
{Uzunov:1993}
D.I. Uzunov, Theory of critical phenomena, World Scientific, Singapore,
 1993.
\bibitem
{Prudnikov:1986} A.P. Prudnikov, Yu.A. Brychkov, and O.I.
Marichev, Integrals and Series, vol. 2, Gordon and Breach, New
York, 1986.
\bibitem
{Stegun:1972}
M. Abramowitz, and I.A. Stegun, Handbook of mathematical functions, Dover
Publ., New York, 1972.

\end{thebibliography}
\end{document}